%% file: arxiv.tex
\documentclass[11pt]{article}

\usepackage{amsmath,amsthm,amssymb,amsfonts,mathtools}
\usepackage[margin=1in]{geometry}
\usepackage{booktabs}
\usepackage{caption}
\usepackage{graphicx}
\usepackage{placeins}
\usepackage{xcolor}
\usepackage[english]{babel}
\usepackage{hyperref}

\hypersetup{
  colorlinks=true,
  linkcolor=blue,
  citecolor=blue,
  urlcolor=blue,
  filecolor=blue,
  pdftitle={Batched and Complete U-Statistics for Trace-Polynomial Estimation from Classical Shadows},
  pdfauthor={Xinyu Song}
}

\theoremstyle{plain}
\newtheorem{theorem}{Theorem}[section]
\newtheorem{proposition}[theorem]{Proposition}
\newtheorem{corollary}[theorem]{Corollary}

\theoremstyle{remark}
\newtheorem{remark}[theorem]{Remark}

\newcommand{\C}{\mathbb C}
\newcommand{\E}{\mathbb E}
\newcommand{\Var}{\operatorname{Var}}
\newcommand{\Cov}{\operatorname{Cov}}
\newcommand{\tr}{\operatorname{tr}}

\newcommand{\norm}[1]{\left\|#1\right\|}
\newcommand{\abs}[1]{\left|#1\right|}
\newcommand{\ket}[1]{\left|#1\right\rangle}
\newcommand{\bra}[1]{\left\langle#1\right|}
\newcommand{\rhohat}{\widehat\rho}
\newcommand{\TB}{\widehat T^{B}}
\newcommand{\TU}{\widehat T^{U}}

\newcommand{\suppsection}[1]{%
  \hyperref[#1]{Supplementary Material, Section~\ref*{#1}}}

\let\arxivtitle\title
\let\arxivauthor\author
\let\arxivdate\date
\let\arxivmaketitle\maketitle
\makeatletter
\let\arxiv@maketitle\@maketitle
\makeatother

\begin{document}

\arxivtitle{Batched and Complete U-Statistics for\\
Trace-Polynomial Estimation from Classical Shadows}
\arxivauthor{Xinyu Song\thanks{E-mail: song.xinyu@mail.shufe.edu.cn.}\\[0.35em]
        \normalsize School of Statistics and Data Science\\
        \normalsize Shanghai University of Finance and Economics}
\arxivdate{July 2026}

\input{main_body.tex}

\input{main_refs.tex}
\clearpage

\setcounter{section}{0}
\setcounter{subsection}{0}
\setcounter{equation}{0}
\setcounter{table}{0}
\setcounter{figure}{0}
\setcounter{footnote}{0}
\setcounter{theorem}{0}
\renewcommand*{\theHsection}{supp.\arabic{section}}
\renewcommand*{\theHsubsection}{supp.\arabic{section}.\arabic{subsection}}
\renewcommand*{\theHequation}{supp.\arabic{equation}}
\renewcommand*{\theHtable}{supp.\arabic{table}}
\renewcommand*{\theHfigure}{supp.\arabic{figure}}

\arxivtitle{Supplementary Material for\\
``Batched and Complete U-Statistics for\\
Trace-Polynomial Estimation from Classical Shadows''}
\arxivauthor{Xinyu Song\\[0.35em]
        \normalsize School of Statistics and Data Science\\
        \normalsize Shanghai University of Finance and Economics}
\arxivdate{July 2026}
\makeatletter
\let\@maketitle\arxiv@maketitle
\makeatother

\input{supplement_body.tex}

\input{supplement_refs.tex}
\end{document}

%% file: main_body.tex
\maketitle

\begin{abstract}
We study estimation of the trace polynomial
$\tr p(P\rho P)$ from global classical shadows, where $\rho$ is an
unknown quantum state and $P$ is a fixed projector. Disjoint batching
and complete U-statistics yield unbiased estimators of the same trace
moments, but assign different sample-size factors to the degenerate
terms in their Hoeffding decompositions. Under the global Clifford
protocol, exact degree-two variance formulas show that, on a null
projected block of rank $s$, the quadratic degenerate term has order
$s^2/N$ under batching and $s^2/N^2$ under complete symmetrization.
For a logarithmic-degree polynomial used in entropy approximation, the
quadratic coefficient raises the batched variance to at least order
$s^2N\log^2N$ at the classical entropy cutoff. For complete
U-statistics, we derive a cross-degree covariance identity and an exact
variance decomposition for polynomial estimators. We also bound every
Hoeffding order at a fixed degree and obtain a growing-dimensional risk
bound for a small-spectrum entropy functional. The higher-order bounds
retain a polynomial dependence on the ambient dimension and therefore
do not cover logarithmically increasing degrees. Monte Carlo
experiments confirm the degree-two formulas, and exact calculations
illustrate the entropy risks.
\end{abstract}



\section{Introduction}

Classical shadows transform randomized quantum measurements into
matrix-valued observations that can be reused to estimate properties
of an unknown state \cite{Aaronson2018,HKP2020}. The resulting theory
is particularly effective for linear observables. If $\rhohat$ denotes
one shadow and $O$ is fixed, then $\tr(O\rhohat)$ is unbiased for
$\tr(O\rho)$, with variance controlled by a shadow norm. Estimation of
purities, R\'enyi entropies, and other polynomial spectral functionals
instead requires products of independent shadows
\cite{BrydgesElben2019,Elben2019,ElbenReview2023,Rath2021,DuEtAl2026}.

Let $P$ be a fixed projector, put $A=P\rho P$, and consider a
polynomial $p(x)=\sum_{k=1}^L a_kx^k$. The statistical problem studied
in this paper is estimation of
\[
  \Phi_p(A):=\tr p(A)=\sum_{k=1}^L a_k\tr(A^k)
\]
from independent global shadows. A product of $k$ shadows provides an
unbiased estimator of $\tr(A^k)$. Such products may be averaged over
disjoint groups of observations or over all distinct $k$-tuples. The
latter construction is the complete U-statistic. The complete
U-statistic is the Rao--Blackwellization of a randomly permuted batched
estimator and consequently has no larger variance for any
square-integrable kernel \cite{Lee1990}. The contribution of this paper
is to determine the size of this variance reduction for
matrix-product kernels under the classical-shadow observation model.

The relevant mechanism is the Hoeffding decomposition
\cite{Hoeffding1948,Lee1990}. For disjoint batching, the variance of
the entire order-$k$ kernel is divided by the number of batches, so
each canonical component receives only a factor of order $N^{-1}$.
For a complete U-statistic, the order-$j$ canonical component receives
a factor of order $N^{-j}$. This distinction becomes important when a
shrinking spectral cutoff produces large polynomial coefficients.

Quantum U-statistics were introduced by Gu\c{t}\u{a} and Butucea
\cite{GutaButucea2010} as operator-valued symmetrizations of
observables measured jointly on several copies. Their results concern
the collective quantum model before measurement and do not apply
directly to the classical matrix-valued observations produced by
shadow protocols. Complete averages are nevertheless widely used in
randomized-measurement estimators of purity and higher moments
\cite{HKP2020,BrydgesElben2019,Elben2019,ZhangSun2021,Rath2021,
ElbenReview2023}. The use of single-copy randomized measurements to
estimate $\tr(\rho^n)$ was considered by van Enk and Beenakker
\cite{VanEnkBeenakker2012}. Incomplete U-statistics have also been
employed in robust classical-shadow estimation of linear observables
\cite{FuKohGohKong2025}. Existing results establish the usefulness of
symmetrization for particular functionals, but do not give a
componentwise comparison between batching and complete averaging for
trace-polynomial kernels.

The degree-two calculation in this paper is exact. Under the global
Clifford protocol, we derive the covariance operator of one projected
shadow and the joint variance of the linear and quadratic trace
statistics. If $A=0$ and $P$ has rank $s$, the quadratic degenerate
component has order $s^2/N$ under batching and $s^2/N^2$ under complete
averaging. These rates include their exact constants. At a null block,
canonical kernels of different degrees are orthogonal even when the
same observations are reused. For the logarithmic-degree Chebyshev
polynomial used below, its quadratic coefficient then gives the lower
bound $s^2L^4/(N\Delta^2)$ for the batched variance. At
$L\asymp\log N$ and $\Delta\asymp(\log N)/N$, this lower bound has
order $s^2N\log^2N$.

For complete U-statistics, we obtain a covariance identity across
polynomial degrees and an exact decomposition of the variance by
Hoeffding order. Every order can be bounded when the polynomial degree
is fixed. The bound contains the factor $(d+1)^{2(j-1)}$ at order $j$,
which restricts its usefulness in high dimension and prevents an
extension to logarithmically increasing degrees. For a quadratic
approximation to $-x\log x$ on $[0,\Delta]$, the resulting estimator of
$-\tr(A\log A)$ has risk
$O\{(s^2+s\log^2N)/N\}$ at $\Delta=N^{-1/2}$. This rate agrees with
the envelope of the linear rule. The quadratic rule improves the
approximation constant and provides an exact nonlinear variance
formula, but it is not asserted to dominate the linear estimator.

The entropy calculation is an application of the trace-polynomial
results. Polynomial approximation near zero is used in classical
large-alphabet entropy estimation
\cite{Paninski2003,JiaoVenkatHan2015,WuYang2016}, while quantum entropy
under collective measurements has a separate sample-complexity theory
\cite{AISW2020,ODonnellWright2016}. Here $P$ is fixed independently of
the shadows. If it is taken to be the population small-spectrum
projector, the result is an oracle statement; estimation of that
projector and of the full von Neumann entropy is not considered.

This study of trace functionals complements work on state estimation
under Pauli and related measurements
\cite{Wang2013,CaiKimWangYuanZhou2016,CaiKimSongWang2021} and the
broader statistical account of quantum tomography in
\cite{WangSong2020}.

The rest of the paper is organized as follows.
Section~\ref{sec:model} introduces the shadow model and the two
estimators. Section~\ref{sec:hoeffding} gives the general Hoeffding
comparison. Sections~\ref{sec:degree-two} and
\ref{sec:log-degree} contain the exact degree-two results and the
logarithmic-degree lower bound. Section~\ref{sec:fixed-degree} studies
complete U-statistics at fixed degree, and
Section~\ref{sec:entropy-application} gives the entropy application.
Numerical results are reported in Section~\ref{sec:numerics}. Proofs
and implementation details are collected in the Supplementary
Material.

\section{Shadow model and trace-moment estimators}
\label{sec:model}

Let $d=2^q$, and let $\rho\in\C^{d\times d}$ be an unknown density
matrix: $\rho=\rho^*$, $\rho\succeq0$, and $\tr(\rho)=1$.
Under the global Clifford protocol, a uniformly random Clifford
unitary $U$ is applied to a fresh copy of $\rho$, followed by a
computational-basis measurement with outcome $b$. One classical shadow
is
\begin{equation}
  \rhohat=(d+1)U^\dagger\ket b\bra bU-I.
\label{eq:shadow}
\end{equation}
The observations $\rhohat_1,\ldots,\rhohat_N$ are independent and
satisfy
\begin{equation}
  \E\rhohat_t=\rho,\qquad
  \Var\{\tr(O\rhohat_t)\}\le3\norm{O}_F^2
\label{eq:shadow-linear}
\end{equation}
for every traceless Hermitian $O$ \cite{HKP2020}.

Let $P$ be a deterministic projector of rank $s$ and define
\[
  A:=P\rho P,\qquad Y_t:=P\rhohat_tP.
\]
Then $\E Y_t=A$. The projector may represent a prescribed physical
subspace, a deterministic target chosen before observing the shadows,
or an oracle spectral block in an application. The main moment results
require only that $P$ is fixed independently of the data.

For Hermitian matrices $Z_1,\ldots,Z_k$ supported on $P$, define the
fully symmetrized trace kernel
\begin{equation*}
 h_k(Z_1,\ldots,Z_k)
 :=\frac1{k!}\sum_{\pi\in\mathfrak S_k}
   \tr(Z_{\pi(1)}\cdots Z_{\pi(k)}).
\end{equation*}
The kernel is real, symmetric, and multilinear. Independence gives
\[
  \E h_k(Y_1,\ldots,Y_k)=\tr(A^k).
\]

Let $B_k=\lfloor N/k\rfloor$. The disjoint-batch estimator and the
complete U-statistic are, respectively,
\begin{align}
 \TB_{k,N}
 &:=\frac1{B_k}\sum_{b=1}^{B_k}
 h_k(Y_{(b-1)k+1},\ldots,Y_{bk}),\notag\\
 \TU_{k,N}
 &:=\binom Nk^{-1}
 \sum_{1\le i_1<\cdots<i_k\le N}
 h_k(Y_{i_1},\ldots,Y_{i_k}).
\label{eq:complete}
\end{align}
For $k=1$ the two definitions agree. Across different degrees, the
same $N$ shadows may be reused. For $1\le L\le N$ and
$p(x)=\sum_{k=1}^La_kx^k$, write
\[
 \widehat\Phi_p^B:=\sum_{k=1}^La_k\TB_{k,N},
 \qquad
 \widehat\Phi_p^U:=\sum_{k=1}^La_k\TU_{k,N}.
\]
Both estimators are unbiased for $\Phi_p(A)$.

\begin{remark}[Computational cost of completeness]
\label{rem:cost}
At the degrees studied here, batching is not forced by computational
cost. By inclusion--exclusion over index coincidences, sums over
distinct tuples reduce to data aggregates: with
$M_r:=\sum_{i=1}^NY_i^r$,
\[
 2\sum_{i<j}\tr(Y_iY_j)=\tr(M_1^2)-\tr(M_2),
\qquad
 \sum_{i,j,l\ \mathrm{distinct}}\tr(Y_iY_jY_l)
 =\tr(M_1^3)-3\tr(M_2M_1)+2\tr(M_3),
\]
so $\TU_{2,N}$ and $\TU_{3,N}$ are computable in one pass over the
data. At a general fixed degree $k$ the same reduction still gives
cost linear in $N$, but coincidence patterns that interleave repeated
indices require tensor-valued aggregates of the form
$\sum_iY_i^{\otimes r}$, whose number and size grow rapidly with $k$
and the block rank $s$; we make no efficiency claim beyond fixed low
degree. The batched estimator is retained as the simplest design and
as a negative benchmark.
\end{remark}

\section{Hoeffding variances of the two estimators}
\label{sec:hoeffding}

Let $h_{j,k}$ be the canonical order-$j$ Hoeffding projection of
$h_k(Y_1,\ldots,Y_k)$ and put
\[
 \zeta_{j,k}:=\E h_{j,k}(Y_1,\ldots,Y_j)^2.
\]
The variance formulas for the two constructions are given below.

\begin{proposition}[Batched and complete Hoeffding variances]
\label{prop:hoeffding-comparison}
For every $1\le k\le N$,
\begin{align}
 \Var(\TB_{k,N})
 &=\frac1{B_k}\sum_{j=1}^k\binom{k}{j}\zeta_{j,k},
\label{eq:batched-hoeffding}\\
 \Var(\TU_{k,N})
 &=\sum_{j=1}^k
   \frac{\binom{k}{j}^2}{\binom Nj}\zeta_{j,k}.
\label{eq:complete-hoeffding}
\end{align}
\end{proposition}

The variance direction in Proposition~\ref{prop:hoeffding-comparison}
also follows directly from Rao--Blackwellization. Let $\Pi$ be an
independent uniform permutation of $\{1,\ldots,N\}$ and form the batched
statistic from $Y_{\Pi(1)},\ldots,Y_{\Pi(N)}$. Conditional on the
observed sample, averaging over $\Pi$ assigns equal weight to every
$k$-subset and therefore gives $\TU_{k,N}$. Exchangeability then yields
\[
 \Var(\TU_{k,N})\le\Var(\TB_{k,N}).
\]
This ordering is standard \cite{Lee1990}. Equations
\eqref{eq:batched-hoeffding} and \eqref{eq:complete-hoeffding} identify
the canonical components and constants that determine the variance
difference.

The first projection has order $N^{-1}$ under either construction. At
order $j\ge2$, however, the complete statistic receives the factor
$N^{-j}$, while batching still provides only $B_k^{-1}\asymp k/N$.
Consequently, an argument that controls only $\zeta_{1,k}$ cannot upper
bound the variance of a batched nonlinear kernel.

For projected shadows satisfying $A\preceq\Delta P$, the
one-shadow inequality gives
\begin{equation}
  \zeta_{1,k}\le3s\Delta^{2(k-1)}.
\label{eq:first-projection}
\end{equation}
The first projection does not control the full kernel variance. At
$k=2$, writing
$E_t:=Y_t-A$ gives
\[
 \tr(Y_1Y_2)-\tr(A^2)
 =\tr(E_1A)+\tr(E_2A)+\tr(E_1E_2).
\]
The final term is the canonical second-order component. It need not
shrink with $\Delta$, even when $A=0$.

\section{Degree-two variance formulas}
\label{sec:degree-two}

Write
\[
 m:=\tr A,\qquad \tau:=\tr(A^2),\qquad E_t:=Y_t-A,
\]
and define
\begin{align*}
 v_0&:=\E\{\tr(E_1)\}^2,&
 v_1&:=\E\{\tr(AE_1)\}^2,\\
 v_2&:=\E\{\tr(E_1E_2)\}^2,&
 c_{01}&:=\E\{\tr(E_1)\tr(AE_1)\}.
\end{align*}
Let
\[
 \widehat T_1:=N^{-1}\sum_{t=1}^N\tr(Y_t),\qquad
 \widehat T_2^B:=B^{-1}\sum_{b=1}^B\tr(Y_{2b-1}Y_{2b}),
 \quad B=\lfloor N/2\rfloor,
\]
and let $\widehat T_2^U$ be defined by
\eqref{eq:complete} with $k=2$.

\begin{theorem}[Exact degree-two variances]
\label{thm:degree-two}
For arbitrary real coefficients $a_1,a_2$,
\begin{align}
 \Var(a_1\widehat T_1+a_2\widehat T_2^B)
 &=\frac{a_1^2v_0}{N}
   +\frac{a_2^2(2v_1+v_2)}{B}
   +\frac{4a_1a_2c_{01}}{N},
\label{eq:degree-two-batched}\\
 \Var(a_1\widehat T_1+a_2\widehat T_2^U)
 &=\frac{a_1^2v_0}{N}
   +a_2^2\left\{\frac{4v_1}{N}
   +\frac{2v_2}{N(N-1)}\right\}
   +\frac{4a_1a_2c_{01}}{N}.
\label{eq:degree-two-complete}
\end{align}
Under the global Clifford protocol,
\begin{align}
 v_0={}&\frac{d+1}{d+2}(s+2m)-m^2
 -\frac{s^2+2sm}{d+2},
\label{eq:v0}\\
 v_1={}&\frac{d+1}{d+2}\{\tau+2\tr(A^3)\}-\tau^2
 -\frac{m^2+2m\tau}{d+2},
\label{eq:v1}\\
 c_{01}={}&\frac{d+1}{d+2}(m+2\tau)-m\tau
 -\frac{sm+m^2+s\tau}{d+2}.
\label{eq:c01}
\end{align}
Moreover, the covariance operator $\mathcal C_A$ of one projected
shadow on the real Hilbert space of $P$-supported Hermitian matrices,
defined by $\langle O,\mathcal C_AR\rangle:=\E\{\tr(OE)\tr(RE)\}$,
acts explicitly as
\begin{equation}
 \mathcal C_A(R)
 =\alpha_d(R+AR+RA)-\tr(AR)\,A
 -\frac{\tr(R)\,(P+A)+\tr(AR)\,P}{d+2},
 \qquad
 \alpha_d:=\frac{d+1}{d+2},
\label{eq:cov-operator}
\end{equation}
and $v_2=\tr(\mathcal C_A^2)$ has the scalar closed form
\begin{align}
 v_2
 ={}&\Bigl(\alpha_d^2+\frac{1}{(d+2)^2}\Bigr)
 (s^2+4sm+2m^2+2s\tau)\notag\\
 &-2\alpha_d\Bigl\{\tau+2\tr(A^3)+\frac{s+4m+4\tau}{d+2}\Bigr\}
 +\tau^2+\frac{2(m^2+2m\tau)}{d+2}.
\label{eq:v2-scalar}
\end{align}
In particular,
\begin{equation}
 v_2
 \le3\mathcal V_{d,s}(A),
 \qquad
 \mathcal V_{d,s}(A)
 :=\frac{(d+1)(s+1)(s+2m)}{d+2}-s-2m-\tau.
\label{eq:v2-bound}
\end{equation}
\end{theorem}

\begin{remark}[Relation to purity estimation]
Randomized-measurement estimators of $\tr(\rho^n)$ and, in particular,
complete pair averages for purity precede the present analysis
\cite{VanEnkBeenakker2012,HKP2020,Elben2019,ElbenReview2023,Rath2021}.
When $P=I$ and $a_1=0$, Theorem~\ref{thm:degree-two} reduces to the
usual first- and second-order Hoeffding structure of a purity
U-statistic. Theorem~\ref{thm:degree-two} additionally treats a fixed
projected block, retains the covariance $c_{01}$ between the linear
and quadratic statistics, and gives both an operator representation
and a scalar formula for $v_2$. These quantities are needed for the
rank-sensitive null-block comparison and the entropy calculation
below.
\end{remark}

The proof uses the projective third-moment identity of the Clifford
orbit \cite{Webb2016}, rather than a fourth-design identity. The same
calculation applies to Haar randomized measurements because their
first three projective moments agree. The proof of
Theorem~\ref{thm:degree-two}, including the derivation of
\eqref{eq:cov-operator} and \eqref{eq:v2-scalar}, is given in
\suppsection{supp:degree-two}.

\begin{corollary}[Sharp null-block comparison]
\label{cor:null-comparison}
Suppose $A=P\rho P=0$ and $1\le s\le d-1$. Then
\begin{equation}
 v_2=(s^2-1)\left(\frac{d+1}{d+2}\right)^2
 +\left(\frac{d+1-s}{d+2}\right)^2
 \ge\frac{s^2}{4}.
\label{eq:null-v2}
\end{equation}
Consequently,
\begin{align}
 \Var(a_1\widehat T_1+a_2\widehat T_2^B)
 &\ge\frac{a_2^2s^2}{2N},
\label{eq:null-batched-lower}\\
 \Var(a_1\widehat T_1+a_2\widehat T_2^U)
 &\ge\frac{a_2^2s^2}{2N(N-1)}.
\label{eq:null-complete-lower}
\end{align}
For the pure quadratic moment, the exact ratio is
\[
 \frac{\Var(\widehat T_2^B)}{\Var(\widehat T_2^U)}
 =\frac{N(N-1)}{2\lfloor N/2\rfloor},
\]
which is of order $N$.
\end{corollary}

The condition $A=0$ is nonempty for every $1\le s\le d-1$: one may
take any density matrix supported on $P^\perp$. It is useful because
the target trace moments vanish while the inverse-measurement noise
inside the projected shadows remains nonzero.

\section{A lower bound for logarithmic-degree batching}
\label{sec:log-degree}

Consider the polynomial sequence used for the entropy application.
Let $g(y)=-y\log y$ on $[0,1]$, with $g(0)=0$, and let
\[
 Q_L^\star(y)=c_0+\sum_{k=1}^Lc_kT_k(2y-1)
\]
be its shifted-Chebyshev truncation. With
$\widetilde g(u):=-(u+1)\log\{(u+1)/2\}/2$, the coefficients are
\[
 c_0=\frac1\pi\int_{-1}^1
 \frac{\widetilde g(u)}{\sqrt{1-u^2}}\,du,\qquad
 c_k=\frac2\pi\int_{-1}^1
 \frac{\widetilde g(u)T_k(u)}{\sqrt{1-u^2}}\,du,\quad k\ge1.
\]
Set
\begin{equation}
 Q_L(y):=Q_L^\star(y)-Q_L^\star(0),\qquad
 p_{L,\Delta}(x):=\Delta Q_L(x/\Delta)-x\log\Delta.
\label{eq:entropy-polynomial}
\end{equation}
Then $p_{L,\Delta}(0)=0$ and
\[
 \sup_{0\le x\le\Delta}
 \abs{-x\log x-p_{L,\Delta}(x)}
 \le C\Delta L^{-2}
\]
by standard polynomial approximation results
\cite{Achieser1956,DeVoreLorentz1993,Trefethen2013}. Write
$p_{L,\Delta}(x)=\sum_{k=1}^La_kx^k$.

\begin{theorem}[Null-block lower bound for the batched polynomial]
\label{thm:full-batched-lower}
Let $P$ have rank $1\le s\le d-1$, suppose $P\rho P=0$, and define
\[
 \widehat\Phi_{L,\Delta}^B
 :=\sum_{k=1}^La_k\TB_{k,N}.
\]
Within each degree the statistic uses disjoint batches, while the same
shadows may be reused arbitrarily across degrees. For $L\ge2$,
\begin{equation}
 a_2=-\frac{(L-1)(L+2)}{3\Delta}.
\label{eq:a2-full}
\end{equation}
If $N\ge2L$, then
\begin{equation}
 \Var(\widehat\Phi_{L,\Delta}^B)
 \ge
 \frac{s^2(L-1)^2(L+2)^2}{18N\Delta^2}.
\label{eq:full-batched-lower}
\end{equation}
Thus, when $L\asymp\log N$ and
$\Delta\asymp(\log N)/N$, the variance is
$\Omega(s^2N\log^2N)$.
\end{theorem}

At a null block every degree-$k$ kernel is canonical in all of its
arguments. Kernels of different degrees are orthogonal even if they
reuse observations: in the product of two terms of different orders,
at least one observation appears in only one kernel, and conditioning
on the remaining observations gives zero. Hence higher-degree terms
cannot cancel the quadratic variance. The coefficient calculation and
the orthogonality argument are given in
\suppsection{supp:log-degree}.

\begin{remark}[Scope]
Theorem~\ref{thm:full-batched-lower} is a variance lower bound for the
raw statistic associated with the particular polynomial
\eqref{eq:entropy-polynomial}. It is not a minimax lower bound, and it
does not rule out other kernels, complete U-statistics, or nonlinear
post-processing such as clipping.
\end{remark}

\section{Complete U-statistics at fixed degree}
\label{sec:fixed-degree}

We next study complete U-statistics beyond degree two. The following
result controls every Hoeffding order and records explicitly the
dimension dependence of the bound.

\begin{theorem}[All Hoeffding orders at fixed degree]
\label{thm:fixed-degree}
Let $1\le L\le N$ and $0\le\Delta\le1$. Suppose
$A=P\rho P\preceq\Delta P$ and write $Y=A+E$. For
$1\le j\le k\le L$, define
\[
 \psi_{j,k}(E_1,\ldots,E_j)
 :=h_k(E_1,\ldots,E_j,A,\ldots,A),
 \qquad
 \zeta_{j,k}:=\E\psi_{j,k}^2,
\]
where the kernel is symmetrized over all $k$ positions. Then
$\psi_{j,k}$ is the canonical order-$j$ Hoeffding projection of
$h_k(Y_1,\ldots,Y_k)$, so that $\zeta_{j,k}$ coincides with the
quantity defined in Section~\ref{sec:hoeffding}, and
\begin{equation}
 \zeta_{j,k}
 \le3s(d+1)^{2(j-1)}\Delta^{2(k-j)}.
\label{eq:fixed-zeta}
\end{equation}
Consequently, for deterministic coefficients $a_1,\ldots,a_L$,
\begin{align}
 \Var\left(\sum_{k=1}^La_k\TU_{k,N}\right)
 \le3s\left[
 \sum_{k=1}^L\abs{a_k}
 \left\{\sum_{j=1}^k
 \frac{\binom{k}{j}^2}{\binom Nj}
 (d+1)^{2(j-1)}\Delta^{2(k-j)}
 \right\}^{1/2}\right]^2.
\label{eq:fixed-polynomial}
\end{align}
\end{theorem}

For every fixed $d$ and $L$, the bound includes every degenerate
component. At higher Hoeffding orders, however, the factor
$(d+1)^{2(j-1)}$ may grow rapidly with the dimension. Thus
Theorem~\ref{thm:fixed-degree} does not establish a useful
dimension-uniform bound when $L\asymp\log N$. Its proof, in
\suppsection{supp:fixed-degree}, iterates the one-shadow inequality
\eqref{eq:shadow-linear} conditionally over the centered arguments.

The canonical projections also determine the covariance between
complete statistics of different degrees, and hence the exact variance
of every polynomial estimator built from them.

\begin{theorem}[Cross-degree covariance and exact polynomial variance]
\label{thm:cross-degree}
In the setting of Theorem~\ref{thm:fixed-degree}, for
$1\le k,\ell\le L$ put
$\zeta_{j,k,\ell}:=\E\{\psi_{j,k}\psi_{j,\ell}\}$ for
$1\le j\le k\wedge\ell$. Then
\begin{equation}
 \Cov(\TU_{k,N},\TU_{\ell,N})
 =\sum_{j=1}^{k\wedge\ell}
 \frac{\binom{k}{j}\binom{\ell}{j}}{\binom Nj}\,
 \zeta_{j,k,\ell},
\label{eq:cross-covariance}
\end{equation}
and for every polynomial $p(x)=\sum_{k=1}^La_kx^k$,
\begin{equation}
 \Var(\widehat\Phi_p^U)
 =\sum_{j=1}^L\binom Nj^{-1}\E\{\Psi_j^2\},
 \qquad
 \Psi_j:=\sum_{k=j}^La_k\binom{k}{j}\psi_{j,k}.
\label{eq:polynomial-variance-identity}
\end{equation}
Moreover
$\abs{\zeta_{j,k,\ell}}\le3s(d+1)^{2(j-1)}\Delta^{k+\ell-2j}$, and
consequently
\begin{equation}
 \Var(\widehat\Phi_p^U)
 \le3s\sum_{j=1}^L
 \frac{(d+1)^{2(j-1)}}{\binom Nj}
 \Bigl\{\sum_{k=j}^L\abs{a_k}\binom{k}{j}\Delta^{k-j}\Bigr\}^2.
\label{eq:polynomial-variance-bound}
\end{equation}
\end{theorem}

By Minkowski's inequality,
\eqref{eq:polynomial-variance-bound} is never larger than
\eqref{eq:fixed-polynomial}: the identity
\eqref{eq:polynomial-variance-identity} keeps distinct Hoeffding
orders exactly orthogonal, and the triangle inequality is applied only
within each order rather than across degrees. The proof is given in
\suppsection{supp:fixed-degree}.

\section{Application to a small-spectrum entropy functional}
\label{sec:entropy-application}

Let $0\preceq A=P\rho P\preceq\Delta P$, where
$0<\Delta\le e^{-1}$, and consider
\[
 H_P(\rho):=-\tr(A\log A).
\]
For $L=2$, the polynomial \eqref{eq:entropy-polynomial} is
\begin{equation}
 p_{2,\Delta}(x)=a_1x+a_2x^2,\qquad
 a_1=\log(1/\Delta)+2\log2-\frac16,\qquad
 a_2=-\frac4{3\Delta}.
\label{eq:quadratic-coefficients}
\end{equation}
Put
\[
 \kappa_2:=
 \sup_{0\le y\le1}
 \abs{-y\log y-\{(2\log2-1/6)y-(4/3)y^2\}}
 \approx0.1324343.
\]
The complete quadratic estimator is
\[
 \widehat H_{P,2}^U:=a_1\widehat T_1+a_2\widehat T_2^U.
\]

\begin{corollary}[Quadratic entropy-functional risk]
\label{cor:entropy-risk}
Let $m=\tr A$. Under the global Clifford protocol,
\begin{align}
 \E\abs{\widehat H_{P,2}^U-H_P(\rho)}^2
 \le{}&\kappa_2^2s^2\Delta^2
 +\frac{C}{N}\{s\log^2(1/\Delta)+s\}\notag\\
 &+\frac{C(s+1)(s+2m)}
 {N(N-1)\Delta^2}.
\label{eq:entropy-risk}
\end{align}
At $\Delta=N^{-1/2}$ and $N\ge4$,
\begin{equation}
 \E\abs{\widehat H_{P,2}^U-H_P(\rho)}^2
 \le C\frac{s^2+s\log^2N}{N}.
\label{eq:entropy-balanced}
\end{equation}
\end{corollary}

The proof is given in \suppsection{supp:entropy}.

For comparison, the linear rule
$\widehat H_{P,1}=\log(1/\Delta)\widehat T_1$ has approximation error
at most $s\Delta/e$ and variance at most
$3s\log^2(1/\Delta)/N$. It therefore has the same leading envelope at
$\Delta=N^{-1/2}$. The quadratic rule improves the certified
approximation constant from $1/e$ to $\kappa_2$, but adds a nonlinear
variance term. Corollary~\ref{cor:entropy-risk} does not assert
uniform risk dominance.

The cross-degree theory of Theorem~\ref{thm:cross-degree} extends the
construction to every fixed degree, at the price of the
dimension-dependent constants and an explicit condition on
$(d,N,\Delta)$.

\begin{corollary}[Fixed-degree entropy rule in growing dimension]
\label{cor:growing-d}
Fix $2\le L\le N$ and $\eta\in(0,1)$, let $p=p_{L,\Delta}$ be the polynomial
\eqref{eq:entropy-polynomial}, and let
$0\preceq A\preceq\Delta P$ with $0<\Delta\le e^{-1}$. If
\begin{equation}
 2L(d+1)^2\le\eta\,N\Delta^2,
\label{eq:growing-d-condition}
\end{equation}
then
\begin{equation}
 \E\abs{\widehat\Phi_p^U-H_P(\rho)}^2
 \le C_{L,\eta}\Bigl\{
 \frac{s^2\Delta^2}{L^4}
 +\frac{s\log^2(1/\Delta)}{N}\Bigr\}.
\label{eq:growing-d-risk}
\end{equation}
\end{corollary}

The proof is given in \suppsection{supp:fixed-degree}.

Condition \eqref{eq:growing-d-condition} requires
$N\Delta^2\to\infty$ and therefore excludes the balanced cutoff
$\Delta=N^{-1/2}$; for cutoffs $\Delta\asymp N^{-\alpha}$ with
$\alpha<1/2$ it permits $d^2=o(N^{1-2\alpha}/L)$. At the balanced
cutoff, only the sharper exact degree-two analysis above yields a useful
dimension-free bound: its degenerate constant $(s+1)(s+2m)$ in
\eqref{eq:entropy-risk} is dimension-free, which the generic factor
$(d+1)^{2(j-1)}$ of Theorem~\ref{thm:cross-degree} does not
reproduce. Whether those dimension factors can be replaced by
rank-sensitive constants at every order remains open.

The regime in Corollary~\ref{cor:growing-d} is nonempty. For example,
fix $L$ and let $\Delta=N^{-\alpha}$, $d=O(N^\beta)$, and $s\le d$,
where
\[
 0<\alpha<\frac12,\qquad
 \beta<\min\left\{\alpha,\frac{1-2\alpha}{2}\right\}.
\]
Then \eqref{eq:growing-d-condition} holds for all sufficiently large
$N$, and both terms on the right-hand side of
\eqref{eq:growing-d-risk} tend to zero, uniformly over
$0\preceq A\preceq\Delta P$ and projectors of rank $s\le d$.

When $P$ is fixed in advance, $\widehat H_{P,2}^U$ is fully
data-based. If $P$ is taken to be the population projector onto
eigenvalues below $\Delta$, then $H_P(\rho)$ is the small-spectrum
contribution to the von Neumann entropy and the result is an oracle
application. Estimating that projector and combining the small block
with a large-spectrum estimator are separate problems not addressed
here.

\section{Numerical results}
\label{sec:numerics}

We report two Monte Carlo experiments for the degree-two formulas and
an exact risk calculation for the entropy application.
Table~\ref{tab:variance} examines the null-block formulas for the
unscaled quadratic moment. The Clifford experiment
uses exact two-qubit global-Clifford sampling of the protocol in
Section~\ref{sec:model}, with $d=4$, $s=1$, and $5000$ Monte Carlo
replicates; the sampler enumerates all $60$ pure two-qubit stabilizer
states and draws from their exact output probabilities. The Haar
experiment uses $d=8$, $s=2$, and $300$ replicates. In both cases
$P\rho P=0$. The complete statistic is evaluated in linear time from
the first identity of Remark~\ref{rem:cost},
\[
 \sum_{i<j}\tr(Y_iY_j)
 =\frac12\left\{
 \tr\left[\left(\sum_iY_i\right)^2\right]
 -\sum_i\tr(Y_i^2)\right\}.
\]

\begin{table}[ht]
\centering
\caption{Null-block variances of the unscaled quadratic moment.
The ratio columns divide empirical by exact variance.}
\label{tab:variance}
\begin{tabular}{cccccc}
\toprule
ensemble & $N$ & $\widehat{\Var}(\widehat T_2^B)$ & ratio$_B$
& $\widehat{\Var}(\widehat T_2^U)$ & ratio$_U$ \\
\midrule
Clifford & $50$  & $1.773\!\times\!10^{-2}$ & $0.998$
& $3.47\!\times\!10^{-4}$ & $0.957$ \\
Clifford & $100$ & $8.799\!\times\!10^{-3}$ & $0.990$
& $9.1\!\times\!10^{-5}$  & $1.012$ \\
Clifford & $200$ & $4.386\!\times\!10^{-3}$ & $0.987$
& $2.2\!\times\!10^{-5}$  & $0.993$ \\
\addlinespace
Haar & $50$  & $1.092\!\times\!10^{-1}$ & $0.935$
& $2.348\!\times\!10^{-3}$ & $0.985$ \\
Haar & $100$ & $5.937\!\times\!10^{-2}$ & $1.017$
& $5.39\!\times\!10^{-4}$  & $0.914$ \\
Haar & $200$ & $2.888\!\times\!10^{-2}$ & $0.989$
& $1.60\!\times\!10^{-4}$  & $1.088$ \\
\bottomrule
\end{tabular}
\end{table}

The ratios in Table~\ref{tab:variance} are close to one. The batched
and complete variances decrease at the predicted rates $N^{-1}$ and
$N^{-2}$, respectively. This experiment concerns the variance
mechanism rather than a complete entropy estimator. Replicate-level
standard errors were not retained. Under a Gaussian fourth-moment
benchmark, the standard error of a variance ratio is approximately
$8.2\%$ for $R=300$ and $2.0\%$ for $R=5000$; the corresponding value
can be larger for a heavy-tailed quadratic kernel. The $4.3\%$
deviation in the first Clifford complete-U row is therefore compatible
with Monte Carlo variation.

Table~\ref{tab:nonnull} examines the formulas in
Theorem~\ref{thm:degree-two} away from the null block, with Monte
Carlo standard errors. The configuration uses exact two-qubit
global-Clifford sampling with $d=4$ and a rank-$2$ block whose
eigenbasis is rotated both globally and inside the block, so that
neither $P$ nor $A$ is aligned with the computational basis. The block
spectrum is $\{0.03,0.06\}$, the remaining mass lies on $P^\perp$, and
$a_1,a_2$ are the coefficients \eqref{eq:quadratic-coefficients} at
$\Delta=0.1$. Each row compares a Monte Carlo estimate based on $5000$
replicates with the exact value: the moments $v_0,v_1,c_{01}$ from
\eqref{eq:v0}--\eqref{eq:c01}; the degenerate moment $v_2$ from the
scalar form \eqref{eq:v2-scalar}, which agrees with the operator
assembly $\tr(\mathcal C_A^2)$ to $10^{-12}$; and the full variances
\eqref{eq:degree-two-batched} and \eqref{eq:degree-two-complete} at
$N\in\{50,200\}$. Every ratio is within about one standard error of
one.

\begin{table}[ht]
\centering
\caption{Non-null Monte Carlo verification of
Theorem~\ref{thm:degree-two} under exact two-qubit global-Clifford
sampling ($d=4$, $s=2$, block spectrum $\{0.03,0.06\}$,
$\Delta=0.1$, $5000$ replicates). The ratio column divides the
empirical value by the exact one; SE is the Monte Carlo standard
error of the ratio.}
\label{tab:nonnull}
\begin{tabular}{ccccc}
\toprule
quantity & $N$ & exact & ratio & SE \\
\midrule
$v_0$ & \text{n/a} & $1.0819$ & $1.001$ & $0.001$ \\
$v_1$ & \text{n/a} & $2.650\!\times\!10^{-3}$ & $1.000$ & $0.001$ \\
$c_{01}$ & \text{n/a} & $4.925\!\times\!10^{-2}$ & $1.001$ & $0.001$ \\
$v_2$ & \text{n/a} & $2.7677$ & $1.001$ & $0.003$ \\
$\Var(a_1\widehat T_1+a_2\widehat T_2^B)$ & $50$ & $19.80$
 & $1.009$ & $0.021$ \\
$\Var(a_1\widehat T_1+a_2\widehat T_2^U)$ & $50$ & $0.5228$
 & $1.003$ & $0.025$ \\
$\Var(a_1\widehat T_1+a_2\widehat T_2^B)$ & $200$ & $4.951$
 & $1.007$ & $0.020$ \\
$\Var(a_1\widehat T_1+a_2\widehat T_2^U)$ & $200$ & $0.05500$
 & $0.995$ & $0.024$ \\
\bottomrule
\end{tabular}
\end{table}

The closed-form variances in Theorem~\ref{thm:degree-two} also permit
exact evaluation of the risks of the entropy rules in
Section~\ref{sec:entropy-application}, without simulation error. We
take $d=16$, a block of rank $s\in\{2,4\}$,
the balanced cutoff $\Delta=N^{-1/2}$, and the spectrum
$\lambda_i=i\Delta/(s+1)$ for $i=1,\ldots,s$; the state is completed
arbitrarily on $P^\perp$, on which the risks do not depend. The
degenerate moment $v_2$ is assembled from the covariance operator
\eqref{eq:cov-operator} on the $s^2$-dimensional block, and the closed
forms \eqref{eq:v0}--\eqref{eq:c01} and \eqref{eq:v2-scalar} are
cross-checked against quadratic forms and the trace of the square of
the assembled operator. Table~\ref{tab:exact-risk} compares
the linear rule $\widehat H_{P,1}=\log(1/\Delta)\widehat T_1$, the
complete quadratic rule $\widehat H_{P,2}^U$, and the batched quadratic
rule $\widehat H_{P,2}^B:=a_1\widehat T_1+a_2\widehat T_2^B$.

\begin{table}[ht]
\centering
\caption{Exact risks for the entropy application at the balanced
cutoff $\Delta=N^{-1/2}$, with $d=16$ and spectrum
$\lambda_i=i\Delta/(s+1)$. All entries are exact evaluations of the
formulas of Theorem~\ref{thm:degree-two}; no Monte Carlo error is
involved.}
\label{tab:exact-risk}
\begin{tabular}{ccccccc}
\toprule
$s$ & $N$ & $\mathrm{bias}^2_{\rm lin}$ & $\mathrm{bias}^2_{\rm quad}$
& $\mathrm{MSE}(\widehat H_{P,1})$
& $\mathrm{MSE}(\widehat H_{P,2}^U)$
& $\mathrm{MSE}(\widehat H_{P,2}^B)$ \\
\midrule
$2$ & $10^3$ & $4.05\!\times\!10^{-4}$ & $2.49\!\times\!10^{-5}$
& $2.09\!\times\!10^{-2}$ & $3.23\!\times\!10^{-2}$
& $1.28\!\times\!10^{1}$ \\
$2$ & $10^4$ & $4.05\!\times\!10^{-5}$ & $2.49\!\times\!10^{-6}$
& $3.61\!\times\!10^{-3}$ & $4.66\!\times\!10^{-3}$
& $1.22\!\times\!10^{1}$ \\
$2$ & $10^5$ & $4.05\!\times\!10^{-6}$ & $2.49\!\times\!10^{-7}$
& $5.58\!\times\!10^{-4}$ & $6.57\!\times\!10^{-4}$
& $1.21\!\times\!10^{1}$ \\
\addlinespace
$4$ & $10^3$ & $1.38\!\times\!10^{-3}$ & $1.12\!\times\!10^{-4}$
& $3.69\!\times\!10^{-2}$ & $8.72\!\times\!10^{-2}$
& $5.26\!\times\!10^{1}$ \\
$4$ & $10^4$ & $1.38\!\times\!10^{-4}$ & $1.12\!\times\!10^{-5}$
& $6.33\!\times\!10^{-3}$ & $1.11\!\times\!10^{-2}$
& $5.04\!\times\!10^{1}$ \\
$4$ & $10^5$ & $1.38\!\times\!10^{-5}$ & $1.12\!\times\!10^{-6}$
& $9.74\!\times\!10^{-4}$ & $1.43\!\times\!10^{-3}$
& $4.97\!\times\!10^{1}$ \\
\bottomrule
\end{tabular}
\end{table}

The quadratic approximation reduces the squared bias in
Table~\ref{tab:exact-risk} by a factor of approximately $12.3$ for
$s=4$ and $16.3$ for $s=2$. These factors do not depend on $N$ for
the specified triangular array. The complete quadratic rule has the
same MSE order as the linear rule because its degenerate term is
$O(s^2/N)$ at the balanced cutoff. The linear rule has the smaller MSE
for the spectra considered here, in agreement with the fact that
Corollary~\ref{cor:entropy-risk} does not establish uniform dominance.
For the batched quadratic rule, the term
$a_2^2v_2/B\asymp v_2/(\Delta^2N)$ is of constant order at
$\Delta=N^{-1/2}$, and its risk stays near $32v_2/9$ across three
decades of $N$. This calculation illustrates at the balanced cutoff
the variance effect established by
Theorem~\ref{thm:full-batched-lower}.

\FloatBarrier

\section{Conclusion}

We studied the effect of tuple averaging on unbiased trace-polynomial
estimators constructed from a common classical-shadow sample. The
Hoeffding decomposition gives an exact explanation of the variance
difference. Disjoint batching assigns an $N^{-1}$ factor to every
canonical component, whereas complete symmetrization assigns an
$N^{-j}$ factor to the component of order $j$. The degree-two formulas
make this difference explicit under the global Clifford protocol,
including the covariance between the linear and quadratic statistics.

For the entropy-motivated polynomial, the null-block lower bound shows
that batching cannot control the variance at the classical cutoff.
Complete U-statistics control every Hoeffding order at a fixed degree
and yield a growing-dimensional risk bound for the small-spectrum
entropy functional. The present higher-order bounds retain polynomial
dependence on the dimension and do not cover logarithmically increasing
degrees. Further progress requires rank-sensitive bounds for products
of centered shadows. Incomplete U-statistics
\cite{Blom1976,ChenKato2019,FuKohGohKong2025} may provide an alternative
when complete aggregation is computationally unattractive. These
questions arise for entropy and other nonsmooth spectral functionals.

%% file: supplement_body.tex
\arxivmaketitle

This supplement contains the proofs and numerical implementation
details for the article. We use the notation of the main text.
In particular, $P$ is a deterministic rank-$s$ projector,
$A=P\rho P$, $Y=P\rhohat P$, and $E=Y-A$.

\section{Unbiasedness and Hoeffding variances}

\subsection{Unbiased trace moments}

The fully symmetrized kernel is
\[
 h_k(Z_1,\ldots,Z_k)
 =\frac1{k!}\sum_{\pi\in\mathfrak S_k}
 \tr(Z_{\pi(1)}\cdots Z_{\pi(k)}).
\]
For independent projected shadows $Y_1,\ldots,Y_k$,
multilinearity and iterated expectation give
\[
 \E\tr(Y_{\pi(1)}\cdots Y_{\pi(k)})
 =\tr\{(\E Y_{\pi(1)})\cdots(\E Y_{\pi(k)})\}
 =\tr(A^k)
\]
for every permutation $\pi$. Thus both $\TB_{k,N}$ and $\TU_{k,N}$
are unbiased for $\tr(A^k)$.

\subsection{Proof of Proposition~\ref{prop:hoeffding-comparison}}

Let
\[
 h_k(Y_1,\ldots,Y_k)-\tr(A^k)
 =\sum_{\varnothing\ne S\subseteq[k]}
 h_{|S|,k}(Y_i:i\in S)
\]
be the canonical Hoeffding decomposition. Orthogonality gives
\begin{equation}
 \Var\{h_k(Y_1,\ldots,Y_k)\}
 =\sum_{j=1}^k\binom{k}{j}\zeta_{j,k}.
\label{supp:eq:kernel-var}
\end{equation}
The $B_k$ batched kernels are independent, so division of
\eqref{supp:eq:kernel-var} by $B_k$ proves
\eqref{eq:batched-hoeffding}.

For the complete U-statistic, a fixed canonical order-$j$ term is
averaged over every $j$-subset of the $N$ observations. Counting how
many $k$-subsets contain a fixed $j$-subset and using orthogonality of
distinct canonical terms gives
\[
 \Var(\TU_{k,N})
 =\sum_{j=1}^k
 \frac{\binom{k}{j}^2}{\binom Nj}\zeta_{j,k},
\]
which is \eqref{eq:complete-hoeffding}. This is the usual complete
U-statistic variance identity \cite{supp-Hoeffding1948,supp-Lee1990}.

\subsection{The first projection}

Because $h_k$ is symmetric and multilinear, the order-one projection
is
\[
 h_{1,k}(Y_1)=\tr\{A^{k-1}(Y_1-A)\}.
\]
If $A\preceq\Delta P$, then
$\norm{A^{k-1}}_F^2\le s\Delta^{2(k-1)}$. The one-shadow inequality
\eqref{eq:shadow-linear} therefore gives
\[
 \zeta_{1,k}
 \le3\norm{A^{k-1}}_F^2
 \le3s\Delta^{2(k-1)},
\]
which proves \eqref{eq:first-projection}.

\section{Degree-two variance calculation}
\label{supp:degree-two}

\subsection{One-shadow covariance operator}

Let $\mathbb H_P$ be the real Hilbert space of Hermitian matrices
supported on $P$, equipped with
$\langle O,R\rangle=\tr(OR)$. Define
\[
 \langle O,\mathcal C_AR\rangle
 :=\E\{\tr(OE)\tr(RE)\},\qquad O,R\in\mathbb H_P.
\]
Put
\[
 \alpha_d:=\frac{d+1}{d+2},\qquad
 r_O:=\tr O,\qquad a_O:=\tr(AO).
\]
The required bilinear form is
\begin{align}
 \langle O,\mathcal C_AR\rangle
 ={}&\alpha_d\{\tr(OR)+\tr[A(OR+RO)]\}-a_Oa_R\notag\\
 &-\frac{r_Or_R+a_Or_R+a_Rr_O}{d+2}.
\label{supp:eq:covariance}
\end{align}

Write $\phi=U^\dagger\ket b$ for the stabilizer state observed before
the inverse measurement channel and
$Q_\phi=\ket\phi\bra\phi$. If $\phi$ is sampled uniformly from the
Clifford orbit, with the multiplicity induced by $(U,b)$, the actual
output law satisfies
\begin{equation}
 \E_{\rho}f(\phi)
 =d\,\E_{\rm unif}\{\langle\phi,\rho\phi\rangle f(\phi)\}.
\label{supp:eq:output-tilt}
\end{equation}
The multiqubit Clifford group is a unitary 3-design
\cite{supp-Webb2016}. Hence, for $t\le3$,
\begin{equation}
 \E_{\rm unif}Q_\phi^{\otimes t}
 =\frac1{d(d+1)\cdots(d+t-1)}
 \sum_{\pi\in\mathfrak S_t}W_\pi,
\label{supp:eq:third-moment}
\end{equation}
where $W_\pi$ is the permutation operator associated with $\pi$.

Since
\[
 \tr(OY)=(d+1)\langle\phi,O\phi\rangle-r_O,
\]
substitution of \eqref{supp:eq:output-tilt} and
\eqref{supp:eq:third-moment} yields
\begin{align*}
 \E\{\tr(OY)\tr(RY)\}
 ={}&\alpha_d\{r_Or_R+a_Or_R+a_Rr_O+\tr(OR)\\
 &\hspace{37mm}+\tr[A(OR+RO)]\}\\
 &-r_Or_R-a_Or_R-a_Rr_O.
\end{align*}
Subtracting $\E\tr(OY)\E\tr(RY)=a_Oa_R$ proves
\eqref{supp:eq:covariance}. The calculation uses projective moments
only through order three.

\subsection{The lower-order terms}

Substitution of $(O,R)=(P,P)$ into
\eqref{supp:eq:covariance} gives
\[
 v_0
 =\frac{d+1}{d+2}(s+2m)-m^2
 -\frac{s^2+2sm}{d+2},
\]
which is \eqref{eq:v0}. The choices $(O,R)=(A,A)$ and $(P,A)$ give
\eqref{eq:v1} and \eqref{eq:c01}, respectively.

The covariance operator is positive semidefinite. For any
$O\in\mathbb H_P$, subtracting
$(\tr O/d)I$ does not change $\tr\{O(\rhohat-\rho)\}$ because both
$\rhohat$ and $\rho$ have trace one. It can only decrease the
Frobenius norm. Therefore \eqref{eq:shadow-linear} implies
\[
 \norm{\mathcal C_A}_{\rm op}\le3.
\]
In particular,
\begin{equation}
 v_0\le3s,\qquad v_1\le3\tr(A^2),\qquad
 \abs{c_{01}}\le\sqrt{v_0v_1}.
\label{supp:eq:linear-bounds}
\end{equation}

\subsection{The fully degenerate term}

Let $\{F_\ell\}_{\ell=1}^{s^2}$ be a real Hilbert--Schmidt
orthonormal basis of $\mathbb H_P$. Independence and Parseval's
identity give
\begin{align*}
 v_2
 &=\sum_{\ell,r}
 \E\{\tr(F_\ell E_1)\tr(F_rE_1)\}
 \E\{\tr(F_\ell E_2)\tr(F_rE_2)\}\\
 &=\tr(\mathcal C_A^2).
\end{align*}
Since $\mathcal C_A\succeq0$ and
$\norm{\mathcal C_A}_{\rm op}\le3$,
\[
 v_2\le3\tr(\mathcal C_A)=3\E\norm{Y-A}_F^2.
\]

Put $q=\langle\phi,P\phi\rangle$. By \eqref{supp:eq:output-tilt} and
\eqref{supp:eq:third-moment} with $t=2$,
\[
 \E q
 =d\,\E_{\rm unif}\{\langle\phi,\rho\phi\rangle\langle\phi,P\phi\rangle\}
 =d\cdot\frac{\tr\rho\,\tr P+\tr(\rho P)}{d(d+1)}
 =\frac{s+m}{d+1},
\]
using $\tr(\rho P)=\tr(P\rho P)=m$. With $t=3$, summing the six
permutation contractions for the triple $(\rho,P,P)$,
\[
 \E q^2
 =d\cdot\frac{s^2+s+2sm+2m}{d(d+1)(d+2)}
 =\frac{(s+1)(s+2m)}{(d+1)(d+2)}.
\]
Because
$Y=(d+1)\ket{P\phi}\bra{P\phi}-P$,
\begin{align*}
 \tr(\mathcal C_A)
 &=\E\norm{Y}_F^2-\norm A_F^2\\
 &=\frac{(d+1)(s+1)(s+2m)}{d+2}-s-2m-\tr(A^2).
\end{align*}
This proves \eqref{eq:v2-bound}.

\subsection{Operator form and a scalar formula for
\texorpdfstring{$v_2$}{v2}}

Collecting the trace pairings of \eqref{supp:eq:covariance} against
$O$ yields the operator identity \eqref{eq:cov-operator}: for all
$O,R\in\mathbb H_P$,
\[
 \langle O,\mathcal C_AR\rangle
 =\tr\Bigl[O\Bigl\{\alpha_d(R+AR+RA)-\tr(AR)A
 -\frac{\tr(R)(P+A)+\tr(AR)P}{d+2}\Bigr\}\Bigr].
\]
Although $\mathcal C_A$ acts on the real Hilbert space $\mathbb H_P$,
we take operator traces on its complexification
\[
 \mathbb H_P\otimes_{\mathbb R}\mathbb C
 \simeq M_s(\mathbb C).
\]
Extend $\mathcal C_A$ complex-linearly; the trace of this extension
equals its real trace. On the complexification write
$\mathcal C_A=G-T$, where
\[
 G:=\alpha_d(\mathrm{Id}+L_A+R_A),\qquad
 T:=|A\rangle\langle A|
 +\frac{|P\rangle\langle P|+|A\rangle\langle P|+|P\rangle\langle A|}{d+2},
\]
$L_A(R):=AR$ and $R_A(R):=RA$ are the multiplication operators on
$M_s(\mathbb C)$, and $|X\rangle\langle Y|$ denotes the rank-one
operator $R\mapsto\tr(YR)X$. Both $G$ and $T$ are self-adjoint on
$M_s(\mathbb C)$ and preserve $\mathbb H_P$, so
$\tr(\mathcal C_A^2)=\tr(G^2)-2\tr(GT)+\tr(T^2)$.

For any Hilbert--Schmidt orthonormal basis $\{F\}$ of $\mathbb H_P$
one has $\sum_FF_{ab}F_{cd}=\delta_{ad}\delta_{bc}$ on the block, and
therefore $\sum_FFXF=\tr(X)P$ for every $X$ supported on $P$. Hence
\[
 \tr(\mathrm{Id})=s^2,\qquad
 \tr(L_A)=\tr(R_A)=sm,\qquad
 \tr(L_A^2)=\tr(R_A^2)=s\tau,\qquad
 \tr(L_AR_A)=m^2,
\]
and expanding $(\mathrm{Id}+L_A+R_A)^2$ gives
\[
 \tr(G^2)=\alpha_d^2\,(s^2+4sm+2m^2+2s\tau).
\]
For the mixed term, self-adjointness of $G$ gives
$\tr(G\,|X\rangle\langle Y|)=\langle Y,GX\rangle$, and
\[
 \langle A,GA\rangle=\alpha_d\{\tau+2\tr(A^3)\},\qquad
 \langle P,GP\rangle=\alpha_d(s+2m),\qquad
 \langle P,GA\rangle=\langle A,GP\rangle=\alpha_d(m+2\tau),
\]
whence
\[
 \tr(GT)=\alpha_d\Bigl\{\tau+2\tr(A^3)+\frac{s+4m+4\tau}{d+2}\Bigr\}.
\]
Finally, $\tr(|X\rangle\langle Y|\cdot|Z\rangle\langle W|)
=\langle Y,Z\rangle\langle W,X\rangle$ gives
\[
 \tr(T^2)=\tau^2+\frac{2(m^2+2m\tau)}{d+2}
 +\frac{s^2+4sm+2m^2+2s\tau}{(d+2)^2}.
\]
Combining the three displays proves \eqref{eq:v2-scalar}. At a
null block, where $m=\tau=\tr(A^3)=0$, the formula reduces to
$\alpha_d^2s^2-2\alpha_ds/(d+2)+s^2/(d+2)^2$, which agrees with the
spectral evaluation \eqref{eq:null-v2} after expanding
$(d+1-s)^2=(d+1)^2-2(d+1)s+s^2$.

\subsection{Batched and complete variances}

The degree-two decomposition is
\[
 \tr(Y_1Y_2)-\tr(A^2)
 =\tr(AE_1)+\tr(AE_2)+\tr(E_1E_2).
\]
The three terms are mutually orthogonal. Consequently,
\[
 \Var(\widehat T_2^B)=\frac{2v_1+v_2}{B}.
\]
Only the two observations in a quadratic batch overlap with
$\widehat T_1$, and conditioning on the other observation gives
\[
 \Cov(\widehat T_1,\widehat T_2^B)=\frac{2c_{01}}N.
\]
These two identities prove \eqref{eq:degree-two-batched}.

For the complete statistic,
\begin{equation*}
 \widehat T_2^U-\tr(A^2)
 =\frac2N\sum_{i=1}^N\tr(AE_i)
 +\binom N2^{-1}\sum_{i<j}\tr(E_iE_j).
\end{equation*}
The two sums are orthogonal. Distinct terms in the second sum are also
orthogonal, including terms that share one index, because conditioning
on that index leaves a centered independent factor. Hence
\[
 \Var(\widehat T_2^U)
 =\frac{4v_1}{N}+\frac{2v_2}{N(N-1)},\qquad
 \Cov(\widehat T_1,\widehat T_2^U)=\frac{2c_{01}}N.
\]
This proves \eqref{eq:degree-two-complete} and completes the proof of
Theorem~\ref{thm:degree-two}.

\subsection{Proof of Corollary~\ref{cor:null-comparison}}

When $A=0$, equation \eqref{supp:eq:covariance} becomes
\[
 \mathcal C_0(O)
 =\alpha_dO-\frac{\tr O}{d+2}P,
 \qquad O\in\mathbb H_P.
\]
Equivalently, $\mathcal C_0$ has eigenvalue $\alpha_d$ on the
$(s^2-1)$-dimensional traceless subspace and eigenvalue
$(d+1-s)/(d+2)$ in the direction $P$. Therefore
\[
 v_2
 =(s^2-1)\left(\frac{d+1}{d+2}\right)^2
 +\left(\frac{d+1-s}{d+2}\right)^2.
\]
For $s=1$, this is $d^2/(d+2)^2\ge1/4$. For $s\ge2$, the traceless
contribution alone is at least $s^2/4$. This proves
\eqref{eq:null-v2}.

At a null block, $v_1=c_{01}=0$. Since $B\le N/2$,
\[
 a_2^2\frac{v_2}{B}\ge\frac{a_2^2s^2}{2N},
\qquad
 a_2^2\frac{2v_2}{N(N-1)}
 \ge\frac{a_2^2s^2}{2N(N-1)}.
\]
The remaining linear variance is nonnegative, proving
\eqref{eq:null-batched-lower} and
\eqref{eq:null-complete-lower}. When $a_1=0$, division of the two
exact expressions gives the ratio in the corollary.

\section{The logarithmic-degree lower bound}
\label{supp:log-degree}

\subsection{The quadratic coefficient}

For $k\ge2$, direct evaluation of the shifted-Chebyshev coefficient of
$g(y)=-y\log y$ gives
\begin{equation}
 c_k=\frac{(-1)^{k+1}}{k(k^2-1)}.
\label{supp:eq:cheb-coeff}
\end{equation}
To verify the identity, put $u=\cos\theta$ and then
$x=\theta/2$. Since
\[
 g\!\left(\frac{1+\cos\theta}{2}\right)
 =-2\cos^2x\log(\cos x)
\]
and
\[
 \log(\cos x)
 =-\log2+\sum_{r=1}^{\infty}
 \frac{(-1)^{r-1}}r\cos(2rx),
 \qquad 0\le x<\frac\pi2,
\]
with convergence in $L^2(0,\pi/2)$, orthogonality on $[0,\pi/2]$ shows
that, for $k\ge2$, the coefficient of $\cos(2kx)$ in
$\cos^2x\log(\cos x)$ is
\[
 \frac{(-1)^{k-1}}{2k}
 +\frac14\left\{
 \frac{(-1)^{k-2}}{k-1}
 +\frac{(-1)^k}{k+1}\right\}
 =\frac{(-1)^k}{2k(k^2-1)}.
\]
The normalization in the definition of $c_k$ multiplies this
coefficient by $-2$, proving \eqref{supp:eq:cheb-coeff}. In particular,
$\sum_{k>L}\abs{c_k}=O(L^{-2})$. The Chebyshev tail and the subtraction
of its value at zero therefore give
\[
 \sup_{0\le y\le1}\abs{g(y)-Q_L(y)}=O(L^{-2}),
\]
which also verifies the approximation bound used in the article.

Taylor expansion at $-1$ gives
\[
 T_k''(-1)=(-1)^k\frac{k^2(k^2-1)}3.
\]
The coefficient of $y^2$ in $T_k(2y-1)$ is therefore
$2(-1)^kk^2(k^2-1)/3$. Combining this with
\eqref{supp:eq:cheb-coeff}, the coefficient of $y^2$ in $Q_L$ is
\[
 -\frac23\sum_{k=2}^Lk
 =-\frac{(L-1)(L+2)}3.
\]
The rescaling
$p_{L,\Delta}(x)=\Delta Q_L(x/\Delta)-x\log\Delta$
proves \eqref{eq:a2-full}.

\subsection{Cross-degree orthogonality at a null block}

Suppose $A=0$. By multilinearity,
$h_k(Y_1,\ldots,Y_k)$ is canonical in every argument. Consider one
term from a degree-$k$ batched statistic and one term from a
degree-$\ell$ statistic, where $k\ne\ell$. Their index sets have
different cardinalities. At least one observation therefore appears
in only one of the two kernels. Conditional on all remaining
observations, the expectation over that observation is zero. Thus the
two terms are uncorrelated. Summation over all terms shows that the
batched statistics of different degrees are pairwise orthogonal,
regardless of how observations are reused across degrees.

\subsection{Proof of Theorem~\ref{thm:full-batched-lower}}

By cross-degree orthogonality,
\[
 \Var(\widehat\Phi_{L,\Delta}^B)
 =\sum_{k=1}^La_k^2\Var(\TB_{k,N})
 \ge a_2^2\Var(\TB_{2,N}).
\]
At the null block,
$\Var(\TB_{2,N})=v_2/B_2$, where
$B_2=\lfloor N/2\rfloor\le N/2$. Using
$v_2\ge s^2/4$ and \eqref{eq:a2-full},
\begin{align*}
 \Var(\widehat\Phi_{L,\Delta}^B)
 &\ge
 \frac{(L-1)^2(L+2)^2}{9\Delta^2}
 \frac{s^2}{4}\frac2N\\
 &=\frac{s^2(L-1)^2(L+2)^2}{18N\Delta^2}.
\end{align*}
This is \eqref{eq:full-batched-lower}.

\section{Complete U-statistics at fixed degree}
\label{supp:fixed-degree}

\subsection{Canonical projections}

Substitute $Y_i=A+E_i$ into the symmetric multilinear kernel. Expansion
over subsets gives
\begin{equation*}
 h_k(Y_1,\ldots,Y_k)-\tr(A^k)
 =\sum_{\varnothing\ne S\subseteq[k]}
 \psi_{|S|,k}(E_i:i\in S).
\end{equation*}
Because $\E E_i=0$, every $\psi_{j,k}$ has conditional mean zero in
each argument and is therefore canonical. Orthogonality and the
complete-U counting identity give
\[
 \Var(\TU_{k,N})
 =\sum_{j=1}^k
 \frac{\binom{k}{j}^2}{\binom Nj}\zeta_{j,k}.
\]

\subsection{Proof of the projection bound}

Condition on $E_1,\ldots,E_{j-1}$. Full symmetrization permits the
order-$j$ projection to be written as
\[
 \psi_{j,k}=\tr(HE_j)
\]
for a Hermitian, $P$-supported matrix $H$ that is the average of
products containing $j-1$ conditioned $E$'s and $k-j$ copies of $A$.
The shadow \eqref{eq:shadow} satisfies
$\norm Y_{\rm op}\le d$, so
$\norm E_{\rm op}\le d+1$. Also
$\norm A_{\rm op}\le\Delta$. On the rank-$s$ block,
\[
 \norm H_F
 \le\sqrt{s}(d+1)^{j-1}\Delta^{k-j}.
\]
Conditionally applying \eqref{eq:shadow-linear} gives
\[
 \E_{E_j}\{\tr(HE_j)\}^2
 \le3\norm H_F^2
 \le3s(d+1)^{2(j-1)}\Delta^{2(k-j)}.
\]
The bound is uniform in the conditioned shadows, so averaging over
them proves \eqref{eq:fixed-zeta}.

Finally, the triangle inequality in $L^2$ gives
\[
 \left\|\sum_{k=1}^La_k
 (\TU_{k,N}-\E\TU_{k,N})\right\|_2
 \le\sum_{k=1}^L\abs{a_k}
 \norm{\TU_{k,N}-\E\TU_{k,N}}_2.
\]
Substituting the complete-U variance identity and
\eqref{eq:fixed-zeta}, then squaring, proves
\eqref{eq:fixed-polynomial} and Theorem~\ref{thm:fixed-degree}.

\subsection{Proof of Theorem~\ref{thm:cross-degree}}

Expanding every kernel of the complete statistic over subsets, a fixed
canonical term $\psi_{j,k}(E_i:i\in S)$ with $\abs S=j$ appears in
exactly $\binom{N-j}{k-j}$ of the $\binom Nk$ index sets, and
$\binom{N-j}{k-j}/\binom Nk=\binom kj/\binom Nj$. Hence
\begin{equation}
 \TU_{k,N}-\tr(A^k)
 =\sum_{j=1}^k\frac{\binom kj}{\binom Nj}
 \sum_{\abs S=j}\psi_{j,k}(E_i:i\in S).
\label{supp:eq:complete-representation}
\end{equation}
Two canonical terms based on distinct index sets are orthogonal: some
observation belongs to exactly one of the two sets, and conditioning
on the remaining observations annihilates that factor. Taking
covariances of the representations for degrees $k$ and $\ell$
therefore retains only common index sets, of size
$j\le k\wedge\ell$, and there are $\binom Nj$ of them:
\[
 \Cov(\TU_{k,N},\TU_{\ell,N})
 =\sum_{j=1}^{k\wedge\ell}
 \frac{\binom kj\binom\ell j}{\binom Nj^2}\,
 \binom Nj\,\zeta_{j,k,\ell},
\]
which is \eqref{eq:cross-covariance}. Summing
\eqref{supp:eq:complete-representation} against the coefficients
$a_k$ and grouping by index set gives
\[
 \widehat\Phi_p^U-\Phi_p(A)
 =\sum_{j=1}^L\binom Nj^{-1}
 \sum_{\abs S=j}\Psi_j(E_i:i\in S),
\]
and orthogonality across index sets proves
\eqref{eq:polynomial-variance-identity}. The Cauchy--Schwarz
inequality and \eqref{eq:fixed-zeta} give
\[
 \abs{\zeta_{j,k,\ell}}
 \le(\zeta_{j,k}\,\zeta_{j,\ell})^{1/2}
 \le3s(d+1)^{2(j-1)}\Delta^{k+\ell-2j},
\]
and the triangle inequality in $L^2$ within each fixed order yields
\[
 \E\{\Psi_j^2\}
 \le3s(d+1)^{2(j-1)}
 \Bigl\{\sum_{k=j}^L\abs{a_k}\binom kj\Delta^{k-j}\Bigr\}^2,
\]
which proves \eqref{eq:polynomial-variance-bound}.
\qed

\subsection{Proof of Corollary~\ref{cor:growing-d}}

By \eqref{eq:entropy-polynomial}, the monomial coefficients of
$p_{L,\Delta}$ are $a_1=q_1+\log(1/\Delta)$ and
$a_k=q_k\Delta^{1-k}$ for $2\le k\le L$, where $q_1,\ldots,q_L$ are
the monomial coefficients of $Q_L$ and depend only on $L$. Hence, for
a constant $c_L$,
\[
 \abs{a_1}\le\log(1/\Delta)+c_L,
 \qquad
 \abs{a_k}\le c_L\Delta^{1-k},\quad2\le k\le L .
\]
Write $S_j:=\sum_{k=j}^L\abs{a_k}\binom kj\Delta^{k-j}$; then
$S_1\le\log(1/\Delta)+c_L'$ and $S_j\le c_L'\Delta^{1-j}$ for
$j\ge2$. Substituting into
\eqref{eq:polynomial-variance-bound}, the order-one term is at
most $C_Ls\log^2(1/\Delta)/N$, using $\log(1/\Delta)\ge1$. For
$j\ge2$, the bound $\binom Nj\ge(N/j)^j\ge(N/L)^j$ gives
\[
 \frac{(d+1)^{2(j-1)}S_j^2}{\binom Nj}
 \le C_L\,\frac LN\,\beta^{j-1},
 \qquad
 \beta:=\frac{(d+1)^2L}{N\Delta^2}\le\frac\eta2<\frac12
\]
under \eqref{eq:growing-d-condition}, and the geometric sum over
$2\le j\le L$ is at most $\beta/(1-\beta)\le\eta$. Hence
\[
 \Var(\widehat\Phi_p^U)
 \le C_{L,\eta}\,\frac{s\{\log^2(1/\Delta)+1\}}N
 \le C_{L,\eta}\,\frac{s\log^2(1/\Delta)}N .
\]
The approximation bias satisfies
$\abs{\Phi_p(A)-H_P(\rho)}\le CsL^{-2}\Delta$ by the uniform bound
following \eqref{eq:entropy-polynomial}. Adding the squared bias
proves \eqref{eq:growing-d-risk}. The requirement $j\le N$ in the
binomial bound holds because \eqref{eq:growing-d-condition}
forces $N\ge2L(d+1)^2/\eta\ge2L$.
\qed

\section{Entropy-functional risk bound}
\label{supp:entropy}

For $0\le x\le\Delta$, rescaling $x=\Delta y$ gives
\[
 -x\log x-p_{2,\Delta}(x)
 =\Delta\left[
 -y\log y-\left\{
 (2\log2-1/6)y-\frac43y^2
 \right\}\right].
\]
Therefore the trace approximation bias over the rank-$s$ block is at
most $\kappa_2s\Delta$.

For the variance in \eqref{eq:degree-two-complete}, the first-order
part can be grouped as
\[
 \frac1N\Var\{a_1\tr(E)+2a_2\tr(AE)\}
 \le\frac{2a_1^2v_0+8a_2^2v_1}{N}.
\]
By \eqref{supp:eq:linear-bounds},
$v_0\le3s$ and
$v_1\le3\tr(A^2)\le3s\Delta^2$. For
$0<\Delta\le e^{-1}$, the coefficients in
\eqref{eq:quadratic-coefficients} consequently give
\[
 \frac{2a_1^2v_0+8a_2^2v_1}{N}
 \le\frac{C}{N}\{s\log^2(1/\Delta)+s\}.
\]
Moreover,
\[
 v_2\le3\mathcal V_{d,s}(A)
 \le3(s+1)(s+2m).
\]
The degenerate term in \eqref{eq:degree-two-complete} is therefore at
most
\[
 \frac{C(s+1)(s+2m)}
 {N(N-1)\Delta^2}.
\]
Adding squared bias proves \eqref{eq:entropy-risk}.

At $\Delta=N^{-1/2}$, $m\le s\Delta$ and $N-1\ge N/2$. For $s\ge1$,
\[
 \frac{(s+1)(s+2m)}
 {N(N-1)\Delta^2}
 \le C\frac{s^2+s}{N}.
\]
The result is trivial when $s=0$. Substitution into
\eqref{eq:entropy-risk} proves \eqref{eq:entropy-balanced}.

\section{Numerical implementation}

\subsection{Monte Carlo null-block experiment}
\label{supp:numerics-mc}

For the Clifford experiment, all $60$ pure two-qubit stabilizer
projectors were enumerated. Given $\rho$, a projector $Q_\phi$ was
sampled from its exact output probability
\[
 d\,\tr(\rho Q_\phi)/60,
\]
and the shadow was formed as $(d+1)Q_\phi-I$. For the Haar experiment,
a Haar unitary was generated and the computational-basis outcome was
sampled from the diagonal of the rotated state, exactly as in the
randomized-measurement protocol. The state $\rho$ was supported on
$P^\perp$ in both experiments.

The batched moment was evaluated from consecutive pairs. The complete
moment used
\[
 \widehat T_2^U
 =\binom N2^{-1}\frac12\left\{
 \tr\left[\left(\sum_{i=1}^NY_i\right)^2\right]
 -\sum_{i=1}^N\tr(Y_i^2)\right\},
\]
which is algebraically identical to pair enumeration but requires only
linear time in $N$ after the matrix sum is formed. Exact variances were
calculated from \eqref{eq:null-v2}. The driver script, the shadow
samplers it imports, and the output files behind
Table~\ref{tab:variance} are included in the \texttt{code/}
directory of the submission package:
\begin{verbatim}
code/compare_batched_complete_t2.py
code/haar_shadow_sampler.py
code/t2_variance_clifford.csv
code/t2_variance.csv
\end{verbatim}

\subsection{Exact risk evaluation}

No sampling is involved in Table~\ref{tab:exact-risk}. The script
assembles the covariance operator $\mathcal C_A$ on the
$s^2$-dimensional space of $P$-supported Hermitian matrices from the
bilinear form \eqref{supp:eq:covariance}, evaluated in the eigenbasis
of $A$, and computes $v_2=\tr(\mathcal C_A^2)$ directly. The closed
forms \eqref{eq:v0}--\eqref{eq:c01} are cross-checked against
the quadratic forms $\langle P,\mathcal C_AP\rangle$,
$\langle A,\mathcal C_AA\rangle$, and
$\langle P,\mathcal C_AA\rangle$ of the assembled matrix; agreement
holds at the $10^{-10}$ level. The mean squared errors then follow from
the exact bias $a_1m+a_2\tau-H_P(\rho)$ and the variance formulas
\eqref{eq:degree-two-batched} and
\eqref{eq:degree-two-complete}. The script and its output are
\begin{verbatim}
code/exact_entropy_risk.py
code/entropy_exact_risk.csv
\end{verbatim}

\subsection{Non-null Monte Carlo verification}

For Table~\ref{tab:nonnull}, the state was constructed from a
seeded Haar-distributed global rotation $U_0$ and an independent
in-block rotation: with the isometry $V$ formed by the first two
columns of $U_0$, the block operator is
$A=R_2\,\mathrm{diag}(0.03,0.06)\,R_2^*$ in the frame of $V$, and the
remaining mass $\mathrm{diag}(0.50,0.41)$ lies on the orthogonal
complement. Shadows were drawn by the exact stabilizer-orbit sampler
of Section~\ref{supp:numerics-mc} and compressed as
$Y_t=V^*\rhohat_tV$. The moments $v_0$, $v_1$, $c_{01}$, and $v_2$
were estimated by empirical means of the per-shadow quantities
$\{\tr(E_t)\}^2$, $\{\tr(AE_t)\}^2$, $\tr(E_t)\tr(AE_t)$, and of
$\{\tr(E_1E_2)\}^2$ over disjoint shadow pairs; their standard errors
are sample standard deviations divided by the square root of the
number of terms. The standard error of an empirical variance over $R$
replicates uses the finite-sample formula
$\Var(\widehat\sigma^2)=\{m_4-(R-3)\widehat\sigma^4/(R-1)\}/R$ with
the empirical fourth central moment $m_4$. The script imports the
samplers from \texttt{compare\_batched\_complete\_t2.py} and the
moment routines from \texttt{exact\_entropy\_risk.py}; it and its
output are
\begin{verbatim}
code/verify_degree_two_nonnull.py
code/t2_nonnull_verification.csv
\end{verbatim}